\documentclass[english,aps, preprint]{revtex4}
\usepackage{graphicx}

\makeatletter

\usepackage{babel}
\makeatother

\begin{document}

\preprint{Preprint}

\title{Density Effects on the Negative Refractive Index of a Split Ring Resonator Metamaterial}

\author{Claudio Amabile \footnote{Dr. Amabile is currently with HS Hospital Service, via Zosimo 13, Rome (Electronic address: c.amabile@hshospitalservice.com) }
}
\affiliation{Laboratorio Nazionale Materiali e Dispositivi per la Microelettronica, Consiglio Nazionale delle Ricerche, Via Olivetti 2, I-20041 Agrate Brianza, Italy}

\author{Enrico Prati}
\email{enrico.prati@cnr.it}
\affiliation{Laboratorio Nazionale Materiali e Dispositivi per la Microelettronica, Consiglio Nazionale delle Ricerche, Via Olivetti 2, I-20041 Agrate Brianza, Italy}

\begin{abstract}
Perfect lensing  and cloaking based on complementary media are possible applications of negative refractive index materials. Metamaterials represent the natural candidates to realize such property by tailoring the effective dielectric permittiviy and magnetic permeability. The fine tuning of $n<0$ of metamaterials is limited by coarse grained inclusions which give discreteness to the electromagnetic parameters as a function of their density. 
We study the negative refractive index of a metamaterial as a function of the linear density of the lattice period at microwave frequency. The negative refractive index is obtained by a waveguide filled with a split ring resonator array in a frequency band below the cut off frequency of the waveguide. The special value of $n=-1$ typical of perfect lensing is maintained by tuning the density on a bandwidth of 1 GHz. A cut-off density above which the metamaterial operates is observed. A second critical density above which the transmission saturates is found close to the Pendry limit. Such results on the density of negative refractive index metamaterials open new paths towards the fine tuning of constitutive parameters at wanted values by means of transformation optics. 
\end{abstract}

\newpage

\maketitle

\section{Introduction}

We study the effect of the lattice density on the negative refractive index of a metamaterial in the proximity of $n=-1$, a range where perfect lenses \cite{Pendry00} and complementary media \cite{Lai09} may be realized.
A metamaterial with negative refractive index has been realized in order to experimentally observe its dependence on the lattice density and to determine the threshold density of its collective excitation. 
$n=-1$ metamaterials provide a viable solution to realize superlensing, for which resolution is not restricted by the diffraction limit.\cite{Pendry00,Fang05,Melville05,Taubner06} A slab left handed material (LHM) would enable to obtain a perfect image of a point source by collecting the contribution of the evanescent components of the spectrum of spatial frequencies, which are amplified in the LHM. Even if the losses in the LHM plate destroy super-resolution, Lagarkov and Kissel showed that the near perfect image is still possible \cite{Lagarkov04}. 
Negative parameters metamaterials, when combined with transformation optics \cite{Schurig06}, can be used to realize complementary media for cloaking. Density grading can be required in order to adapt impedance from vacuum to the negative parameters medium like in positive metamaterial cloaking\cite{Pendry09}.

According to classical dielectrics theory, the refractive index of a macroscopic medium $n$ and the constitutive constant tensors $\epsilon$ and $\mu$ depend on both the features of the elementary cell - the single resonator in the case of a metamaterial - and their density. \cite{Jackson} 
In the past, tuning the parameters of the cell has been preferred instead of varying the density.
Negative refractive index has been previously reported \cite{Guven06,Ozbay07} for some metamaterials but its dependence on the lattice density was not investigated. The tuning of the refractive index of a metamaterial by means of lattice density grading has been shown for positive refractive index transformation media cloaking.\cite{Schurig06}
The collective behavior of the system has been analyzed by some authors \cite{Varadan_JAP06,Powell07}.   
The split ring resonator (SRR) based metamaterials have been solved analytically by some authors \cite{Pendry_MTT99,Markos02,Gorkunov_PRE06}. The results disagree on the dependence upon the density of SRRs, in particular those along the direction of their axis. 
The dependence of the electromagnetic constants upon the resonators density was previously demonstrated in some experimental studies \cite{Varadan_JAP06,Powell07}.

The lattice constant dependence of a negative refractive index (NRI) metamaterial is studied. 
For sake of simplicity the negative permeability was realized by means of split ring resonators  while the negative effective permittivity was provided by a sub cut-off square waveguide.  The combination of the two characteristics determines a sub cut-off band where both the effective constitutive constants relevant for the propagation are negative. Such system is described by uniaxial permeability and refractive index tensors.\cite{Hrabar_AP05}
We investigated the lowest density at which the metamaterial behavior holds.
We show that the resonance frequency of the effective medium depends only on the single resonator properties, whereas its bandwidth depends on the density. We provide the dependence of the negative refractive index of the effective medium on the lattice density, as a function of the microwave frequency. $n=-1$ is observed over a bandwidth of 1 GHz by varying the lattice density. Finally, we demonstrate the existence of two threshold densities below which the negative permeability behavior disappears (cut-off density) and the transmitted power and the absorption become density independent (critical density, corresponding to the Pendry limit).  

The paper is organized as follows: in Section II, the theoretical framework of uniaxial metamaterial rectangular waveguides is recalled; in Section III the experimental results are shown: in the first subsection the details of the sample and the setup are exposed; in the second subsection the sub cut-off frequency transmission band is characterized and the frequency dependence of the negative refractive index shown; in the third and fourth subsection the threshold and the saturation are studied; the fifth provides the details of the method used to determine the negative refractive index. In the Conclusion the findings and the implications of the present study on the realization of a $n=-1$ metamaterial for perfect lensing and grading for complementary negative media are listed. 

\section{Theory of SRR based uniaxial metamaterials}

Uniaxial media are a special case of the most general family of gyroelectromagnetic media, described by hermitian tensorial constitutive constants \cite{Prati03,Prati04}
\begin{equation}
\begin{array}{clrr} %
  \hat{\epsilon} = \left(
  \begin{array}{clrr} %
    \epsilon_{x} & 0 & 0 \\
	  0 & \epsilon_{1} & -i \epsilon_{2} \\
	   0 & i \epsilon_{2} & \epsilon_{1} 
  \end{array}
  \right)
&
  \hat{\mu} = \left(
  \begin{array}{clrr} %
    \mu_{x} & 0 & 0 \\ 
  	0 & \mu_{1} & -i \mu_{2} \\
	  0 & i \mu_{2} & \mu_{1} 
  \end{array}
  \right)
\end{array} 
\label{eq:epsilonmu}
\end{equation}
 
A waveguide filled with a split ring resonator lattice consists of a uniaxial medium which is obtained by setting $\epsilon_1=\epsilon_x$, $\epsilon_2=0$ and $\mu_{2}=0$.

The theory of the propagation in a rectangular waveguide filled with SRRs has been reported by Hrabar and co-workers. \cite{Hrabar_AP05}. A rectangular waveguide with a base of side $a$ along the $\hat{x}$ longitudinal direction and a height $b$ along the $\hat{y}$ direction is considered. The propagation occurs along the $\hat{z}$ direction of the waveguide. 
Both the permittivity and permeability of SRRs are anisotropic. In such a case the tensors of the constitutive constants take the form

\begin{equation}
\begin{array}{clrr} %
  \hat{\epsilon} = \epsilon_r \left(
  \begin{array}{clrr} %
   1 & 0 & 0 \\
	0 & 1 & 0 \\
	 0 & 0 & 1 
  \end{array}
  \right)
&
  \hat{\mu} = \left(
  \begin{array}{clrr} %
    \mu_{x} & 0 & 0 \\ 
  	0 & 1 & 0 \\
	  0 & 0 & 1 
  \end{array}
  \right)
\end{array} 
\label{eq:epsilonmuSRR}
\end{equation}
where $\mu_1 =1$ because the resonance of SRRs affects only the component of the permeability tensor ortogonal to the SRR plane and the magnetic response of SRRs along the other directions is negligible. Both $\epsilon_r$ and $\mu_x$ may vary according to the lattice period when a resonance of the SRR array is excited.

For the sake of simplicity, in the following \textit{permeability} refers to the only transverse $\hat{x}$ component. 
The dispersion relation for the $m$-th mode ($m=1,2,3,...$) with $k_x=m \pi / a$ is
\begin{equation}
k_z=\pm \sqrt{\epsilon_r \mu_{x} \left( k^2_0 -\frac{k^2_x}{\epsilon_r \mu_{z}} \right)} 
\end{equation}

The cut-off frequency $\nu_c$  of the $m-$th mode is 
\begin{equation}
\nu_c  = \frac{\nu_{c0}}{\sqrt{\epsilon_r \mu_{z}'}}  
\label{eq:fff}
\end{equation}
where $\nu_{c0}=mc/2a$. Defining $\mu_{x}'=Re[\mu_{x}]$, the dispersion relation for the eigenmodes of the system becomes \cite{Hrabar_AP05}:
\begin{equation}
k_z = \pm k_0\sqrt{\epsilon_r \mu_{x}' \left[ 1-\left( \frac{\nu_c}{\nu} \right)^2 \right]}
\label{eq:k}
\end{equation}

The sign of the wave number, which determines whether the mode is propagating or evanescent, is related to the sign of the electromagnetic constants. When $\epsilon_r >0$ and $\mu_{x}'>0$, which is the standard case, the wave number is positive for $\nu>\nu_c$ (propagating wave) and negative elsewhere (evanescent wave). When $\epsilon_r >0$ and $\mu_{x}'<0$, which is the case of interest, the mode is evanescent for $\nu>\nu_c$ and propagating below the cut-off frequency.

\section{Characterization of the negative refractive index band}

The experiment was carried in a guided microwave circuit including a square waveguide exploited below its cut-off frequency. The negative effective dielectric permittivity of such waveguide, combined with a metamaterial with a suitable negative effective permeability component, has the property to allow propagation below the cut off frequency. The advantage of the double negative waveguide-metamaterials combination with respect to those metamaterials made of two interleaved lattices is the lack of interaction between the inclusions used to switch the sign of the permittivity and the permeability separately. In this section the samples and the experiments are described.

\subsection{Realization of the metamaterial square waveguide}
\begin{figure}[hbt]
\begin{center}\includegraphics[scale=.5, angle=0]{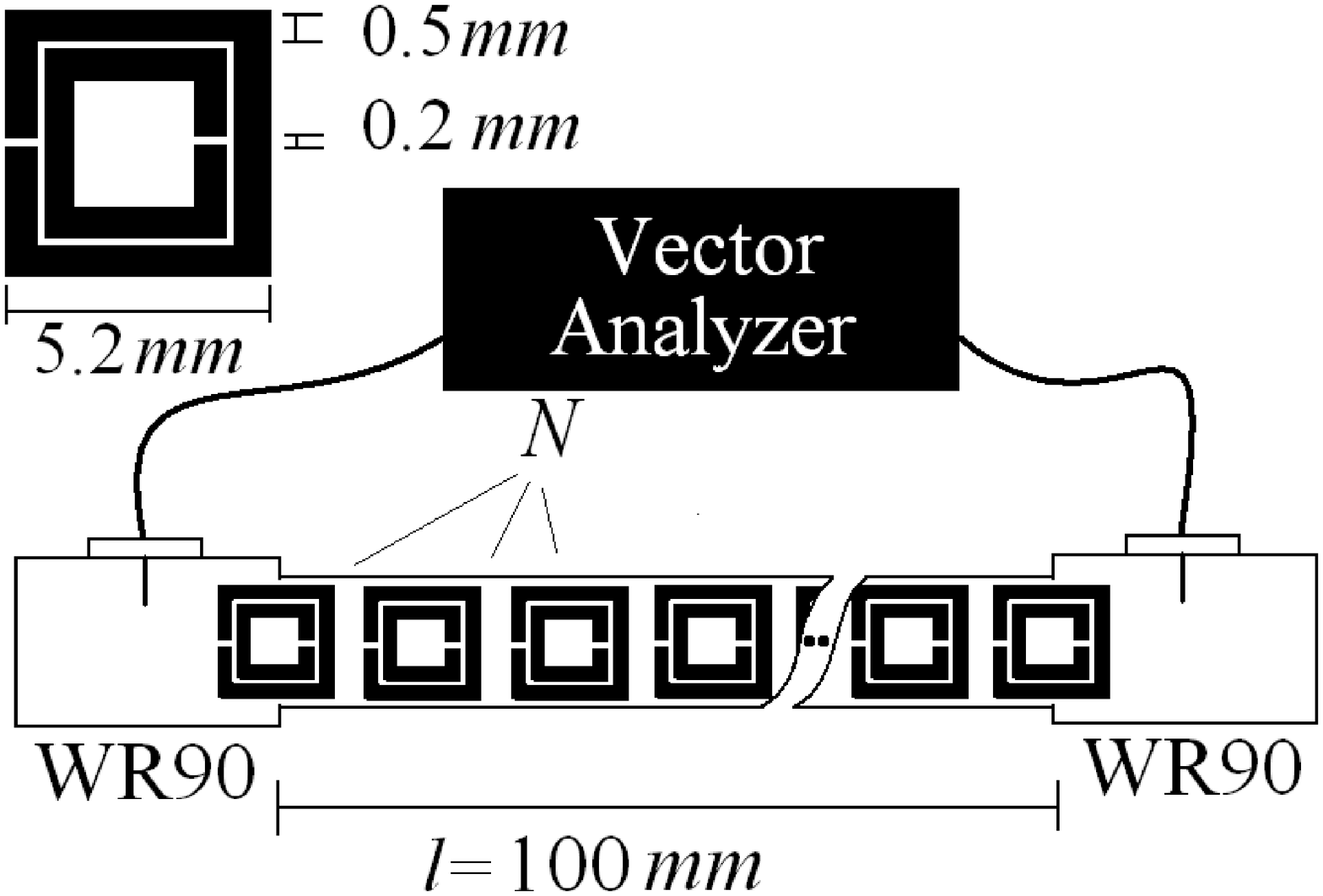}
\caption{The experimental equipment consists of a 2 port vector network analyzer connected to a waveguide system which includes the metamaterial. It is constituted by two flange-to-coaxial line transitions, and an uniaxial metamaterial formed by an array of split ring resonaters periodically spaced along $z$ axis inserted in a specially machined square waveguide.}
\label{fig:setup}
\end{center}
\end{figure}

Eleven samples with different SRRs density were fabricated. Each sample consists of a 10 cm long FR4 slab over which an array of copper SRRs was etched (Figure \ref{fig:setup}). The height of the slab is 6.9 mm, equal to the waveguide height. The number of SRRs varies from sample to sample, ranging from $N=9$ to $N=19$. The distance between the first and last SRR was kept constant in order to have the same electromagnetic length\cite{Chen_PRE04}. The SRRs are square with an external side of $c=5.2$ mm. The gaps between the two rings and between the arms of the same ring are 0.2 mm. The strip is 0.5 mm wide. The substrate is 1.6 mm thick FR4, whose permittivity is 4.2 and the tangent loss is 0.008 at 3 GHz. The Pendry resonance calculated for such parameters of the SRRs is expected at around 9 GHz.\cite{Pendry_MTT99} 
The samples were placed in a square waveguide whose side is 6.9 mm and length $l=$10 cm. The theoretical cut-off frequency of such a waveguide is 21.7 GHz. The slab was placed in the center of the waveguide with the SRRs orthogonal to the H field of the TE10 mode. The gap of the SRRs was also perpendicular to the E field of the TE10 mode. The square waveguide was fed by two WR90 waveguides, one for each end. In a WR90 waveguide only the TE10 mode propagates at frequencies between 8.1 GHz and 12 GHz. Such waveguide carries the TE10 mode in the frequency range of the resonance of the SRRs. The scattering parameters of the waveguide are measured by means of a vector network analyzer. The coaxial line was calibrated by means of TRL method up to the coaxial waveguide transitions. The boundary SRRs are inserted in the waveguide only for a half in order to couple to the waveguide-to-coaxial transitions. The linear density of a $N$ elements sample is $1/ l_N$ where $l_N=l/(N-1)$ is the lattice constant (the distance between the centers of two neighboring SRRs) and $l$ is the length of the waveguide.

\subsection{NRI as a function of the lattice density}
\begin{figure}[hbt]
\begin{center}\includegraphics[scale=.75, angle=0]{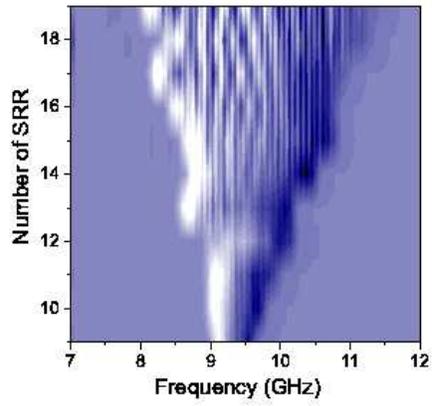}
\caption{The color scale represents the first derivative of $S_{21}$ with respect to microwave frequency $\nu$. The bandwidth increases by increasing the density of SRRs, while the resonance frequency remains fixed at 9.25 GHz.}
\label{fig:zoom}
\end{center}
\end{figure}

Sub cut-off transmission band due to the presence of the SRR medium was shown by Marques  \cite{Marques_PRL02} and Hrabar \cite{Hrabar_AP05}. They also showed that the wave propagation in the sub cut-off band is backward, according to theoretical predictions for NRI media. We observe an analogue transmission band in our experiment. The module of the measured $S_{21}$ is represented in Figure \ref{fig:zoom}. 

The (sub cut-off) frequency of the center of the transmission band is at $\nu_{res}= 9.25$ GHz and it is independent on the density. A narrow bandwidth corresponds to the lowest density. The bandwidth enlarges as the density is increased. This result supports the hypothesis that the overall resonance frequency only depends upon the single resonator properties. The bandwidth enlarges as the density is increased, demonstrating its dependence on the collective properties of the lattice of resonators.

\begin{figure}[hbt]
\begin{center}\includegraphics[scale=.66, angle=0]{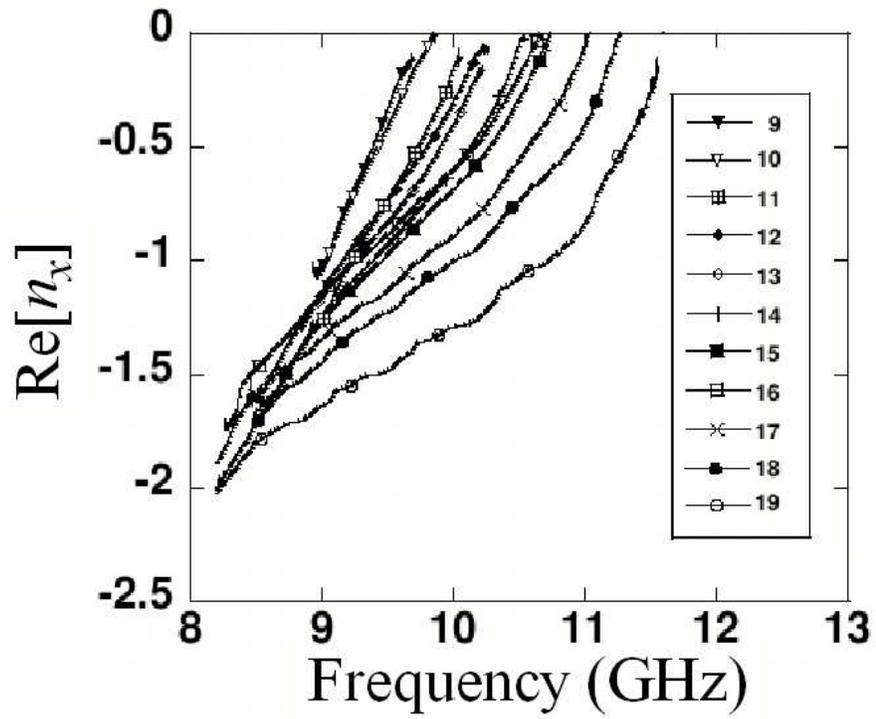}
\caption{Real part of the refractive index as a function of frequency for densities of the lattice ranging from $N=9$ to $N=19$. }
\label{fig:n_final}
\end{center}
\end{figure}

The method used to retrieve the refractive index is discussed in the last subsection of this paragraph.
In Figure \ref{fig:n_final} the negative refractive index $n_x'$ is shown as a function of the frequency $\nu$ for all the available densities. The negative refractive index spans between $0$ and $-2$. By properly adjusting the lattice density the refractive index can be tuned at $n_x=-1$ on a bandwidth of about 1 GHz. Its modulus $\left|n_x \right|$ increases as a function of the density.

\begin{figure}[hbt]
\begin{center}\includegraphics[scale=.75, angle=0]{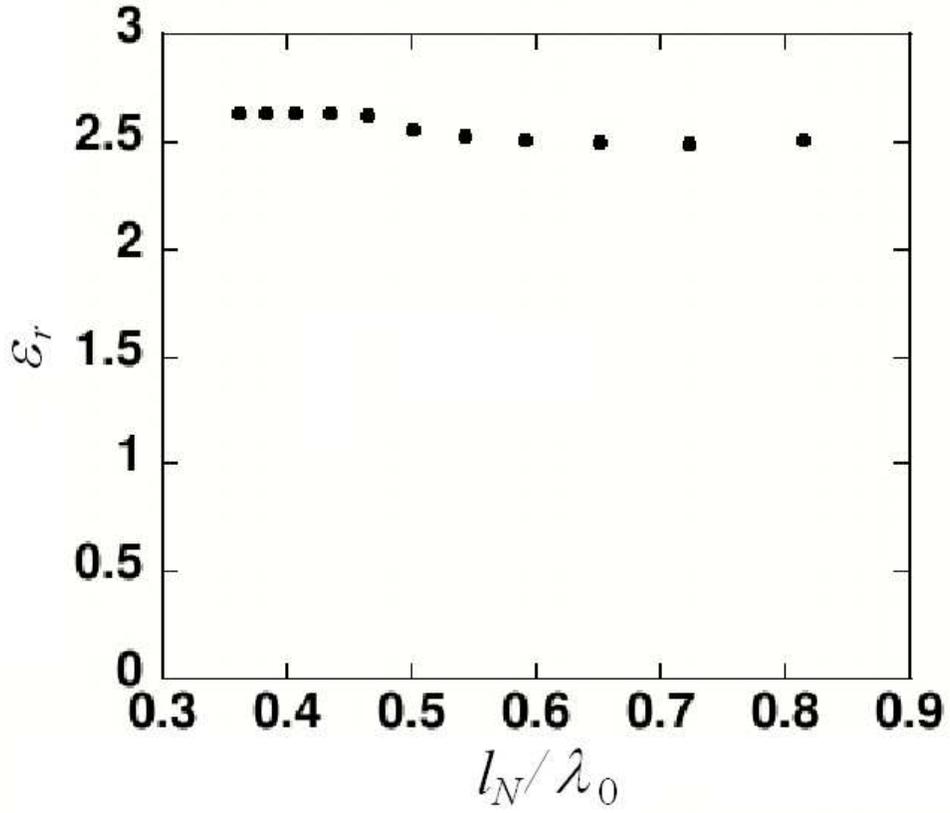}
\caption{The relative dielectric permittivity $\epsilon_r$ calculated from the cut-off frequency of the square waveguide (Eq. \ref{eq:fff}) as a function of the ratio between the lattice constant and the wavelength. The small dependence of $\epsilon_r$ at around 2.5-2.6 on the density, combined with the strong density dependence of $n_x'$ shown in Figure \ref{fig:n_final} implies that the density principally affects the magnetic permeability $\mu_x'$.}
\label{fig:epsilon}
\end{center}
\end{figure}

\begin{figure}[hbt]
\begin{center}\includegraphics[scale=.66, angle=0]{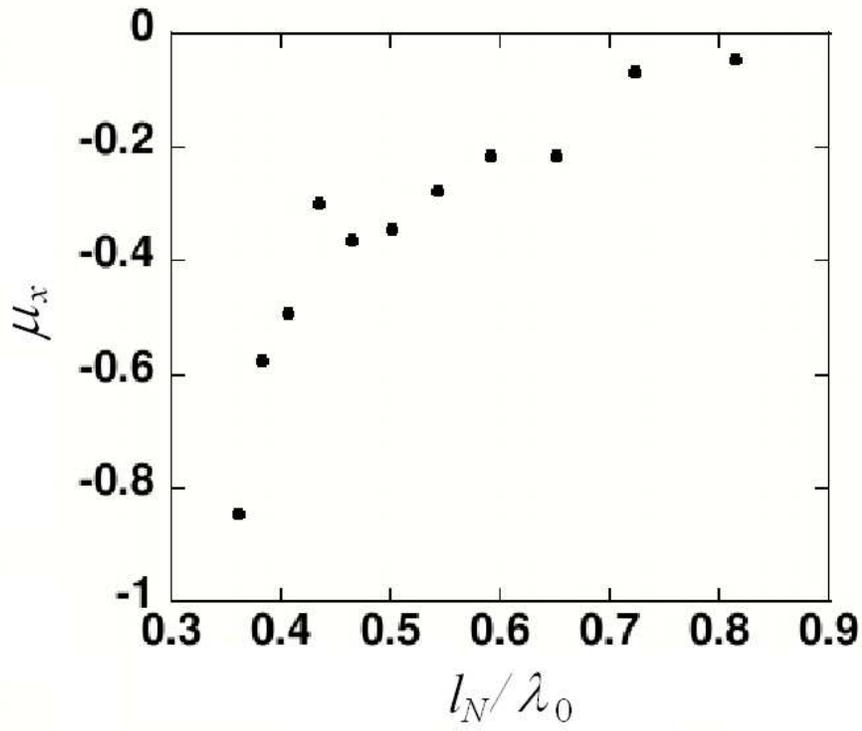}
\caption{Effective permeability $\mu_{x}$ at the arbitrary frequency of 9.5 GHz as a function of the ratio $l_N / \lambda_0$ between the lattice constant and the wavelength. }
\label{fig:permeability}
\end{center}
\end{figure}

The effective relative permittivity $\epsilon_r$ and permeability $\mu_x$ at $\nu_0=9.5$ GHz are shown respectively in Figure \ref{fig:epsilon} and Figure \ref{fig:permeability} as a function of the ratio of the lattice constant $l_N$ over the wavelength in the waveguide  $\lambda_0=c/(\sqrt(\nu_c^2-\nu^2))$. The permittivity $\epsilon_r$ is calculated for each sample as the ratio of the cut-off frequency of the empty waveguide over the forward propagation cut-off frequency of the waveguide loaded with the SRRs. The value retrieved of $\epsilon_r \approx2.5-2.6$ is nearly independent upon the number of SRR, demonstrating that the electric response of the SRR is negligible.

\subsection{Cut-off and critical densities in the NRI band}

\begin{figure}[hbt]
\begin{center}\includegraphics[scale=.75, angle=0]{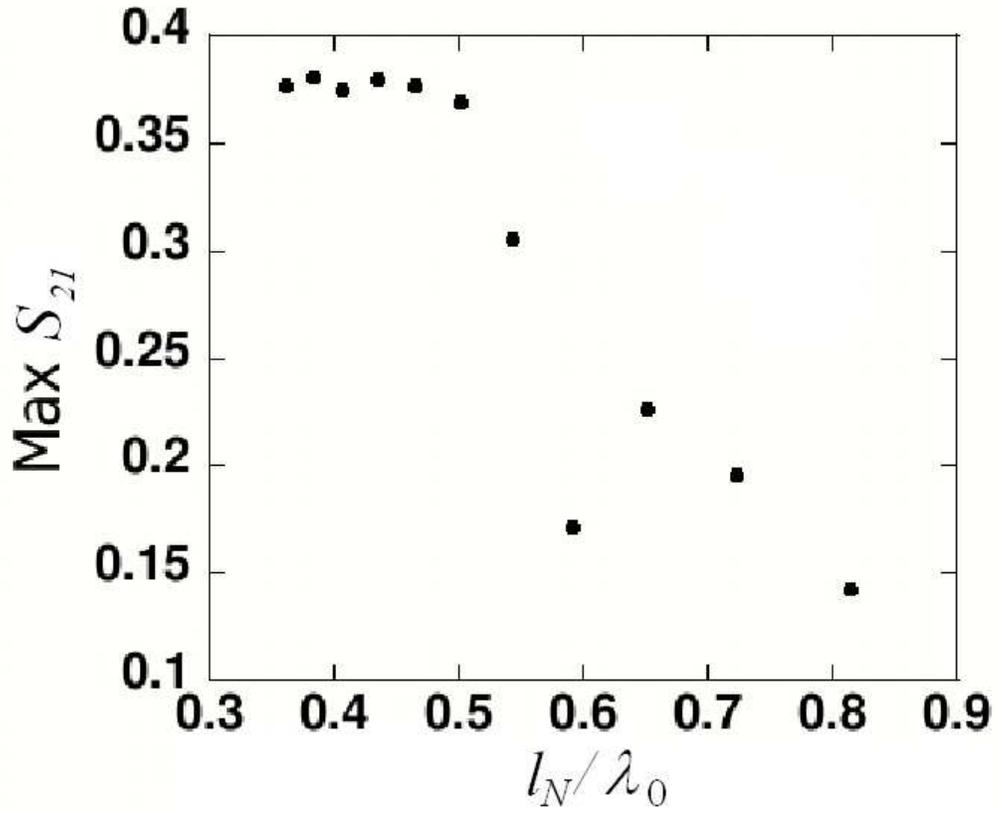}
\caption{Amplitude of the scattering coefficient $S_{21}$ as a function of the ratio between the lattice period and the wavelength. At $l_N / \lambda_0 \approx 0.5$ ($N=14$) a critical linear density $1/l_N$ is found, above which $S_{21}$ is density independent.}
\label{fig:amp}
\end{center}
\end{figure}

\begin{figure}[hbt]
\begin{center}\includegraphics[scale=.75, angle=0]{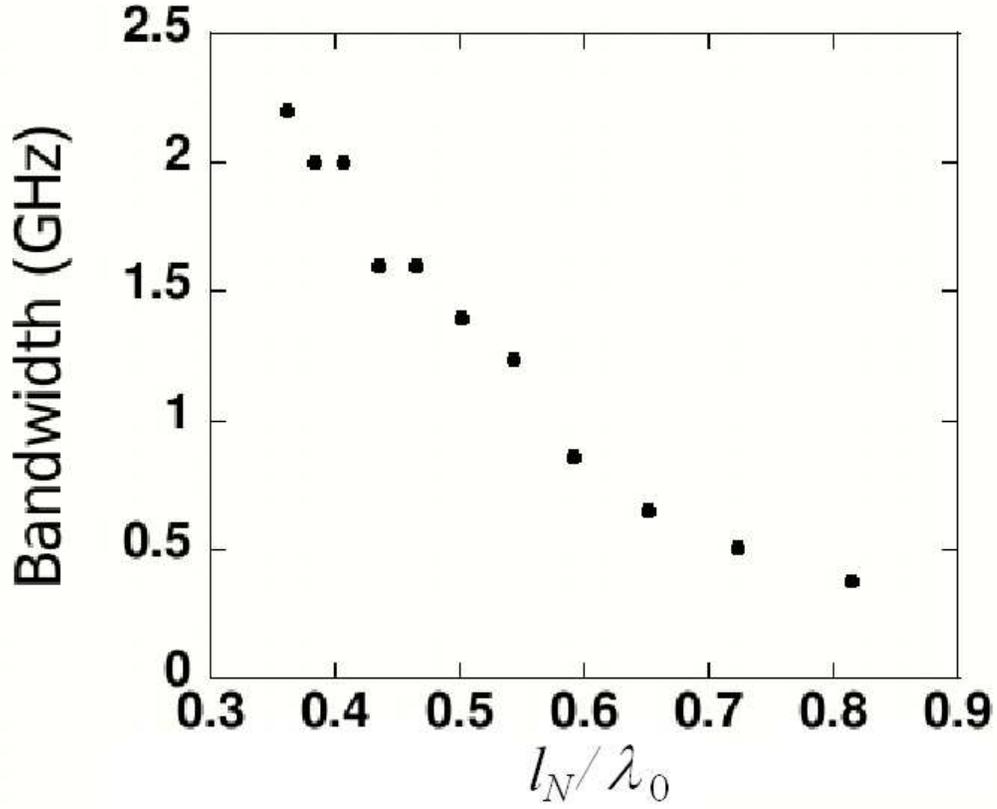}
\caption{Bandwidth of the scattering coefficient $S_{21}$ as a function of the ratio between the lattice period and the wavelength. The vanishing of the sub cut-off band can be ascribed to a reduction of the amplitude of resonance of the permeability, caused by the reduction of the resonators density. As a consequence, the negative permeability region narrows and disappears. Since negative permeability is necessary for sub cut-off propagation, the reduction of the negative permeability region makes the sub cut-off transmission band narrower. 
}
\label{fig:bw}
\end{center}
\end{figure}

The amplitude and bandwidth of the sub cut-off transmission band were measured as a function of the density. The results are shown in Figure \ref{fig:amp} and \ref{fig:bw} respectively. The quantity reported in Figure \ref{fig:amp} is the maximum of the corresponding curve of Figure \ref{fig:bw}. The bandwidth is calculated with the -3 dB criterion with respect to the maximum of the curve.

The strong dependence of the bandwidth on the density indicates that it is a collective property of the metamaterial. Therefore, when the bandwidth of metamaterials based on different resonators is compared \cite{Ran_PRB04}, the density should be considered.
Both the amplitude and the bandwidth of the transmission band tend to zero at around $N=7$ SRRs in the array. The distance between neighboring resonators is $1.67$ cm. The corresponding density defines the threshold cut-off density to be overcome in order to establish a metamaterial regime. 

We conclude that the bandwidth is determined collectively by the lattice. When the bandwidths of metamaterials based on different resonators are compared, as in Ref. \cite{Ran_PRB04}, the density of the sample should be therefore considered.

When the ratio between the lattice period $l_N$ and the wavelength $\lambda=c/\nu$ is lower than 1/4 ($N>14$ for which $l_N / \lambda_0 \leq 0.5$), the amplitude of the transmitted wave flattens as a function of the lattice density. This novel finding suggests the existence of a second threshold density, called \textit{critical }density in the following.
The ratio between such lattice period and the wavelength $\lambda$ at the center of the transmission band at around 9.25 GHz experimentally retrived is close to the Pendry limit of $1/6$. 

Once the NRI regime is achieved, the amplitude of the transmitted wave is almost constant, depending only on the impedance of the effective medium. Such finding on the amplitude suggests that, while some effects of the metamaterial nature of the system arise at a certain density, the metamaterial regime is saturated only above a critical density.

\subsection{Dependence of the losses on the density}

\begin{figure}[hbt]
\begin{center}\includegraphics[scale=.75, angle=0]{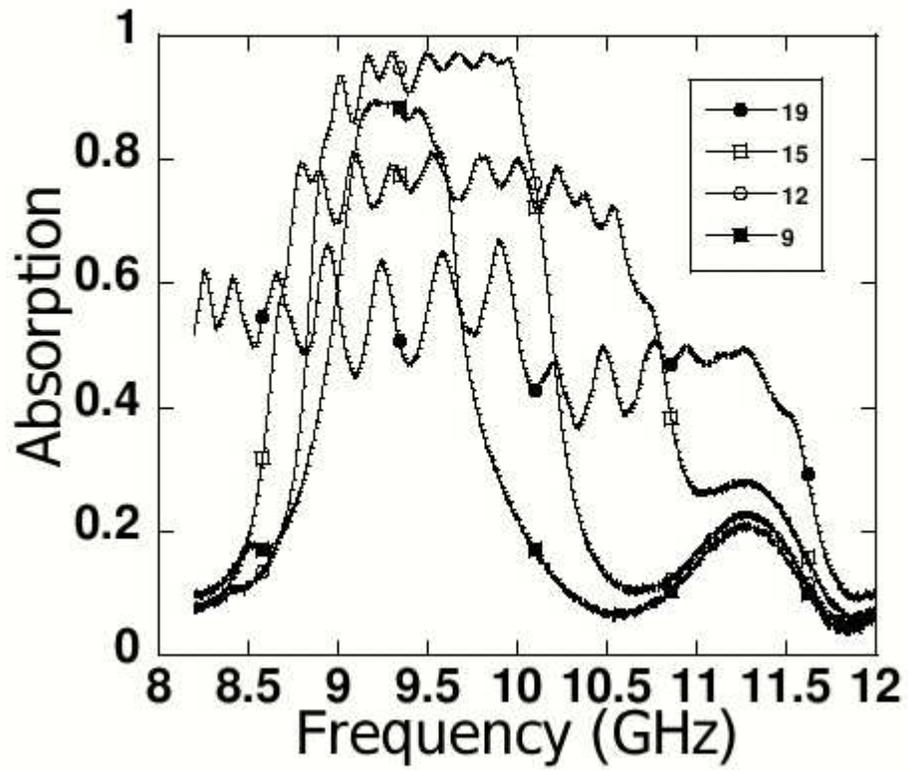}
\caption{Absorbed power $\eta$  as a function of the frequency. For the sake of clarity the absorption of four samples only are shown. }
\label{fig:abs}
\end{center}
\end{figure}

\begin{figure}[hbt]
\begin{center}\includegraphics[scale=.75, angle=0]{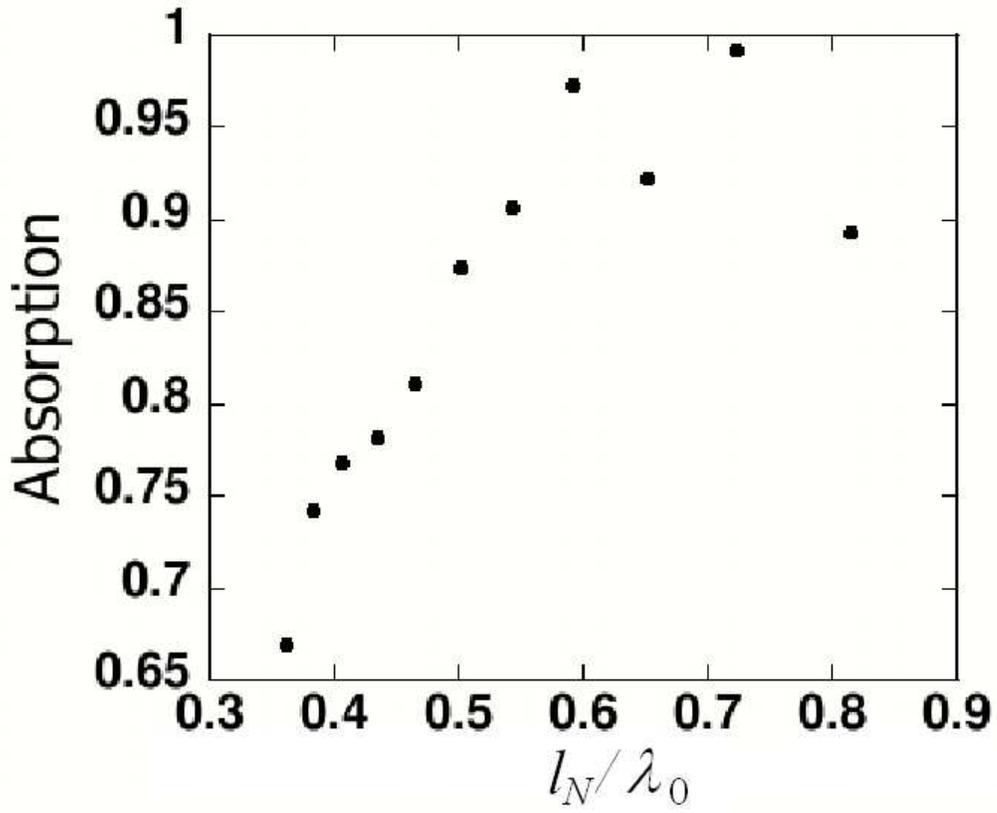}
\caption{Maximum absorbed power $\eta$  as a function of the ratio between the lattice period and the wavelength. At low density the forward wave which is not transmitted to the second port is mainly absorbed instead of being reflected.}
\label{fig:maxabs}
\end{center}
\end{figure}

The ratio between the absorbed over the incident power is retrived from $\eta =1-|S_{11}|^2-|S_{21}|^2$. Its frequency dependence is illustrated in Figure \ref{fig:abs}. The absorbed power is nearly constant in the transmission band. Noticeably $\eta$ is close to 1 for low density samples, while it lowers when the density is increased. To quantify this behavior we plot in Figure \ref{fig:maxabs} the maximum of $\eta$ versus the number of SRRs of the sample. Therefore, at low density the forward wave which is not transmitted to the second port is mainly absorbed instead of being reflected.

Almost all the transmitted power is absorbed when the ratio between the lattice period and the wavelength is higher than 1/4. This number is the same retrieved from the analysis of the transmitted power of Figure  \ref{fig:amp}. 

The metamaterial transmission loss at low density is around -130 dB/m while it is around -40 dB/m in the best case above the critical density. Even though -40 dB/m is considerably high, it is nevertheless orders of magnitude smaller than those below the critical density.

The presence of such critical density for transmitted power absorption should be taken into account when designing resonator based metamaterials. Another interesting feature is that the absorbed power decreases when increasing the resonators density also when the threshold density is overcome. This behavior could be somehow counterintuitive because one could expect that, once the NRI regime is established, the increase of the density of resonators could produce an increase of conduction losses over the metal of the resonators.

Such high transmission losses can be attributed to a high dispersion of the refractive index, which lowers when the density is increased, according to classical electromagnetism. 

\subsection{Determination of the negative refractive index value in the sub cut-off regime}

The coupling between the $TE_{10}$ mode of the feeding waveguide with the mode propagating in the metamaterial can not be determined. Therefore the scattering matrix of the metamaterial can not be worked out and consequently the permittivity and permeability can not be extracted by means of usual reflection/transmission methods \cite{NIST}. 

Nevertheless the real part of the refractive index is extracted directly by the phase of the transmission coefficient under the hypothesis that a single mode is propagating through the square waveguide. Such hypothesis is supported by the observation that only the $TE_{10}$ mode is carried by the WR90 waveguide, which couples mainly to the $TE_{10}$ mode of the square waveguide. Eventual higher order modes generated at the discontinuity at the interface between the metamaterial and the coaxial-waveguide transition do not match the wavelenght prescription for the effective response with respect to the lattice spacing. The $TE_{01}$ mode is ruled out by the selectivity on the field components operated by the anisotropy of the SRRs and by the negligible coupling with the $TE_{10}$ excitation of the feed waveguide.   
The phase of the transmission coefficient for a monomodal wave is
\begin{equation}
\phi = \sum_{i} k_i d_i + 2 \pi \ell
\label{eq:phi}
\end{equation}
where the index $i$ spans over all the media crossed by the wave, $k_i, d_i$ are the wave constant and thickness of the $i$-th medium and $\ell$ is an integer. The measured phase is the sum of the phase shift accumulated in the metamaterial plus the phase shift due to the two coaxial to waveguide transitions. The latter can be accurately evaluated by measuring the phase shift of the line made of the two transitions only. 

From Eq. \ref{eq:phi} the wave number of the wave propagating in the metamaterial was retrieved. By means of equation \ref{eq:k} we obtained the quantity $n_{x}'=-\sqrt{\epsilon_r \mu_{x}'}$, from the wave number. Analogously to the behavior of the refractive index of conventional dielectrics close to their resonance frequencies, the $2\pi \ell$ phase uncertainity of the transmission coefficient is addressed as the refractive index goes to zero at the high-frequency end of the sub cut-off transmission band. Once the $2\pi$ uncertainity is solved at high frequencies, it is wiped away at low frequency by continuity.

\section{Conclusion}
 
A negative refractive index metamaterial based on split ring resonators has been realized and characterized in the microwave frequency domain. The metamaterial has double negative parameters. The sign of the permittivity and of a transverse component of the permeabiltiy are inverted by exploiting a waveguide below its cut-off frequency and a periodic lattice of split ring resonators respectively. The lattice constant dependence of the negative refractive index and of the pass band characteristics have been studied. The modulus of the negative refractive index at fixed frequency is reduced by decresasing the density of the metamaterial. A cut-off density above which the propagation occurs is observed. A further critical density is found, close to the Pendry limit, above which the transmission is stable. The problem of coarse grained discretization of the refractive index due to the small number of cells can be circumvented by exploiting the frequency dependence of the refractive index in a band of around 1 GHz. The density of the cells of the lattice should be taken into account to exploit negative parameters grading for advanced metamaterial for cloaking and perfect lensing.


\begin{acknowledgments}
The authors acknowledge A. Zito (Università di Pisa) for the phase locked experimental equipment. 
\end{acknowledgments}

\newpage

\end{document}